\documentclass[reprint,aps,prb,amsmath,amssymb]{revtex4-2}

\usepackage[]{graphicx}
\usepackage{times}
\usepackage{bm}
\usepackage[ulem=normalem]{changes}
\usepackage[normalem]{ulem}		
\usepackage{color}
\usepackage{xr} 				
\usepackage{cancel}
\usepackage{hyperref}
\usepackage{siunitx}
\usepackage{babel}
\usepackage{subcaption}
\usepackage{mathtools}
\usepackage{dsfont}
\usepackage{tikz}
\usetikzlibrary{decorations.pathmorphing}
\usetikzlibrary{arrows.meta}
\usetikzlibrary{patterns}
\usetikzlibrary{calc,positioning}

\definechangesauthor[color=blue]{SM}

\usepackage{cleveref}
\usepackage{braket}
\DeclareUnicodeCharacter{2212}{\ensuremath{-}}
\DeclareUnicodeCharacter{00A0}{~}

\begin{document}

\title{An \textit{ab initio} based compact analytical formula for inelastic x-ray scattering from spatially localized excitons}

\author{Joris Sturm$^{1}$, Duc Anh Truong$^{1}$, Andrei Benediktovitch$^{2}$, Nina Rohringer$^{2,3}$  and Andreas Knorr$^{1}$}
\affiliation{$^1$ Technische Universität Berlin, Institut für Physik und Astronomie, Nichtlineare Optik und Quantenelektronik, 10623 Berlin, Germany}
\affiliation{$^2$ Deutsches Elektronen-Synchrotron DESY, Hamburg 22607, Germany}
\affiliation{$^3$ Department of Physics, Universität Hamburg, Hamburg 22607, Germany}

\begin{abstract}
In this contribution we provide a compact model for the description of inelastic x-ray scattering of spatially localized excitons. We combine our analytical theory with first principles calculations and apply it to the material lithium fluoride (LiF). A comparison of two different approaches for the underlying single-particle wave functions to calculate the spectrum yield a qualitatively good agreement with experimental data but also shows that a detailed description of the wave functions is essential to obtain qualitative results.
\end{abstract}

\maketitle

\section{Introduction}
Understanding the collective and elementary excitations of condensed matter is crucial for the description of their optical and electronic properties. Among these excitations, excitons, which are composite quasiparticles consisting of Coulomb-bound electrons and holes, play a central role for a wide range of phenomena such as optical absorption or energy transport~\cite{akselrod_visualization_2014, xiao_excitons_2017}. 
Excitons are often categorized depending on their extension in real-space into Wannier, Frenkel or charge transfer (CT) excitons. While Wannier excitons are spatially delocalized over many unit cells of the crystal, Frenkel and CT excitons exhibit a much stronger spatial localization. Frenkel excitons are characterized by an electron and a hole that are predominantly localized at the same atomic or molecular site. In contrast, CT excitons are often considered an intermediate type, in which the electron and hole are localized at different, typically neighboring atomic sites. CT excitons can be found, for example, in alkali halides like LiF, which serves as an exemplary material in this contribution. 
\\
 An established method for the experimental study of excitons is inelastic x-ray scattering (IXS)~\cite{abbamonte_dynamical_2008, balut_quantum_2025, caliebe_dynamic_2000}. Here, an incident x-ray transfers both, energy and momentum, to a material which can be analyzed in the resulting momentum-transfer resolved energy-loss spectrum. In contrast to optical spectroscopy, where the momentum transfer is typically negligible, IXS enables the spectrum to be explored over a finite momentum range. This additional degree of freedom provides exemplarily access to the dispersion of elementary excitations as well as their internal charge distribution after optical pumping and thus offers information that are not obtainable in purely optical measurements. Recent examples include the theoretical description of inelastic XUV scattering to detect Landau orbits~\cite{meyer_proposal_2026} and IXS from spatially delocalized Wannier excitons in two-dimensional (2D) semiconductors~\cite{sturm_theory_2026}.
 In this contribution, we extend the theory for the description of inelastic scattering from spatially localized excitons. In particular, our goal is to provide a compact analytical formula which built a bridge between \textit{ab initio} methods and a microscopic model.   
 \\
To describe IXS from excitons we model the following situation:
 An incoming x-ray $\boldsymbol{A}_{X}$ with wave vector $\boldsymbol{k}_{X}$ and frequency $\omega_{X}$ hits the sample and is inelastically scattered from excitons in the material. The wave vector after scattering is denoted by $\boldsymbol{k}$ and the frequency by $\omega$. 
Inelastic x-ray scattering set-ups measure the momentum and energy resolved (angular resolved) scattered x-ray radiation.
\section{Theoretical framework}
\label{sec:theory_localized_exciton}
 The frequency and momentum resolved stationary spectrum $S(\Delta\boldsymbol{q},\omega)$ of scattered radiation is given by~\cite{vogel_quantum_2006} (cf. Eq. 21 in~\cite{sturm_theory_2026}):
\begin{equation}
\label{eq:definition_spectrum_with_time_ordered_correlation}
S(\Delta \boldsymbol{q},\omega) = \int_{0}^{\infty} d\tau e^{\!-i\omega  \tau} e^{ -\Gamma \tau}  G(\Delta \boldsymbol{q},\tau) +c.c.
\end{equation}
where $\omega$ is the detected frequency, $\Delta\boldsymbol{q}=\boldsymbol{k}_{X}-\boldsymbol{k}$ is the momentum transfer, defined as the difference of the momentum of the incoming x-ray $\boldsymbol{k}_{X}$ and the momentum $\boldsymbol{k}$ of the scattered radiation. $\Gamma$ denotes the width of a spectral filter and
$G(\Delta \boldsymbol{q},\tau) = \lim_{t\to \infty} G(\Delta \boldsymbol{q},t+\tau,t)$ is the steady-state limit of the time-ordered field-field correlation function~\cite{sturm_theory_2026}:
\begin{widetext}
    \begin{equation}
\label{eq:time-ordered-correlation-function ausgeschrieben}
\begin{aligned}
    G(\Delta \boldsymbol{q},t+\tau,t)= \left|\boldsymbol{A}^{\perp}_{X} \omega_{X}\frac{ r_{0}}{ R }\right|^2 \sum_{\substack{\lambda,\lambda' \\ \mu,\mu' \\ \boldsymbol{k_1}, \boldsymbol{k_2} }}& \braket{ a^{\dagger \lambda}_{\boldsymbol{k_1}}(t+\tau) a^{\lambda'}_{\boldsymbol{k_1}+\Delta \boldsymbol{q}}(t+\tau) a^{\dagger \mu}_{\boldsymbol{k_2}+\Delta \boldsymbol{q}}(t)a^{\mu'}_{\boldsymbol{k_2}}(t) } e^{i\omega_{X} \tau}
      \\
      \times& \braket{\lambda,\boldsymbol{k_1}|\lambda', \boldsymbol{k_1}+\Delta \boldsymbol{q}}  \braket{\mu,\boldsymbol{k_2}+\Delta\boldsymbol{q}|\mu', \boldsymbol{k_2}}.
    \end{aligned}
\end{equation}
\end{widetext}
Here, $\boldsymbol{A}^{\perp}_{X}$ is the transverse part of the amplitude and $\omega_{X}$ the frequency of the incoming x-ray, $r_{0}=\frac{ \mu_0 e^2 }{4\pi m }$ is the Thomson scattering length~\cite{alsnielsen_elements_2011} and $R$ is the distance between the target and the detector. The electrons are described by Bloch states where the fermionic annihilation (creation) operators are given by $a^{(\dagger) \lambda}_{\boldsymbol{k}}$ with the band index $\lambda$, electron momentum $\boldsymbol{k}$ and the associated wave functions are Bloch functions $\Psi^{\lambda}_{\boldsymbol{k}}(\boldsymbol{r})=\frac{1}{\sqrt{V}}u^{\lambda}_{\boldsymbol{k}}(\boldsymbol{r}) e^{i \boldsymbol{k}\cdot \boldsymbol{r}}$, which are normalized to the volume of the crystal $V$. The lattice periodic functions $u^{\lambda}_{\boldsymbol{k}}(\boldsymbol{r})$ are also referred to as Bloch factors. In Eq.~\ref{eq:time-ordered-correlation-function ausgeschrieben} we introduced an abbreviation for the overlap of the Bloch factors:
\begin{equation}
\label{eq:definition_Bloch_factor_overlap}
    \braket{\lambda,\boldsymbol{k_1}|\lambda', \boldsymbol{k_1}+\Delta \boldsymbol{q}} = \frac{1}{V_{\text{UC}}} \int_{\text{UC}}d^3r' u^{*\, \lambda}_{\boldsymbol{k_1}}(\boldsymbol{r}') u^{\lambda'}_{\boldsymbol{k_1}+\Delta \boldsymbol{q}}(\boldsymbol{r}'),
\end{equation}
where $V_{\text{UC}}$ is the volume of the unit cell. 
For a description of spatially localized excitons, located at atomic sites $\boldsymbol{R}_{l}$, we apply a basis of spatially localized Wannier functions~\cite{ashcroft_solid_1976,kittel_introduction_2005} for $u^{\lambda}_{\boldsymbol{k}}(\boldsymbol{r})$:
\begin{equation}
    u^{\lambda}_{\boldsymbol{k}}(\boldsymbol{r}) = \frac{1}{\sqrt{N}} \sum_{\boldsymbol{R}_{l}} e^{-i\boldsymbol{k}\cdot(\boldsymbol{r}-\boldsymbol{R}_{l})} w_{\lambda}(\boldsymbol{r}-\boldsymbol{R}_{l}).
\end{equation}
Here, $N$ is the number of unit cells with volume $V_{\text{UC}}$, $\boldsymbol{R}_{l}$ is a lattice vector of the crystal lattice in real-space and $w_{\lambda}(\boldsymbol{r}-\boldsymbol{R}_{l})$ are the Wannier functions of the band $\lambda$. 
In this basis, Eq.~\ref{eq:definition_Bloch_factor_overlap} reads:
\begin{widetext}
    \begin{equation}
    \begin{aligned}
            \braket{\lambda,\boldsymbol{k_1}|\lambda', \boldsymbol{k_1}+\Delta \boldsymbol{q}}   &=\frac{1}{N} \sum_{\boldsymbol{R}_{l},\boldsymbol{R}_{m}}e^{i\boldsymbol{k_1}\cdot(\boldsymbol{R}_{m}-\boldsymbol{R}_{l})} \frac{1}{V_{\text{UC}}}\int_{\text{UC}}d^3r'  w^{*}_{\lambda}(\boldsymbol{r'}-\boldsymbol{R}_{l}) w_{\lambda'}(\boldsymbol{r'}-\boldsymbol{R}_{m}) e^{-i\Delta\boldsymbol{q}\cdot(\boldsymbol{r'}-\boldsymbol{R}_{m})}
    \end{aligned}
\end{equation}
\end{widetext}
Using $\boldsymbol{R}_{n}=\boldsymbol{R}_{m}-\boldsymbol{R}_{l}$, we can rewrite the overlap of the Bloch factors to:
\begin{equation}
    \begin{aligned}
            \braket{\lambda,\boldsymbol{k_1}|\lambda', \boldsymbol{k_1}+\Delta \boldsymbol{q}}             
            &=\sum_{\boldsymbol{R}_{n}}  t^{\lambda,\lambda'}_{\Delta\boldsymbol{q}}(\boldsymbol{R}_{n}) e^{i\boldsymbol{k_1}  \cdot\boldsymbol{R}_{n}}  
    \end{aligned}
\end{equation}
where we introduced:
\begin{equation}
\label{eq:overlap_integral_definition_wannier_function}
    t^{\lambda,\lambda'}_{\Delta\boldsymbol{q}}\!(\boldsymbol{R}_{n}) = \frac{1}{V}\!\int \!d^3r'   w^{*}_{\lambda}(\boldsymbol{r'}) w_{\lambda'}(\boldsymbol{r'}\!-\boldsymbol{R}_{n}) e^{-i\Delta\boldsymbol{q}\cdot(\boldsymbol{r'}\!-\boldsymbol{R}_{n})}.
\end{equation}
Since we consider Coulomb interacting electrons in Eqs.~\ref{eq:definition_Bloch_factor_overlap}-~\ref{eq:overlap_integral_definition_wannier_function} exciton formation is included in the two-time correlation function $\braket{ a^{\dagger \lambda}_{\boldsymbol{k_1}}(t+\tau) a^{\lambda'}_{\boldsymbol{k_1}+\Delta \boldsymbol{q}}(t+\tau) a^{\dagger \mu}_{\boldsymbol{k_2}+\Delta \boldsymbol{q}}(t)a^{\mu'}_{\boldsymbol{k_2}}(t) }$~\cite{sturm_theory_2026}. 
Therefore, we use the Bethe-Salpeter equation (BSE)~\cite{salpeter_relativistic_1951, rohlfing_electron-hole_2000, chen_ab_2019, puschnig_optical_2002,  haber_maximally_2023, quintela_theoretical_2022}:
\begin{equation}
    \label{eq:BSE_eigenvalue_equation}
    \sum_{v' c' \boldsymbol{k'}} H^{e}_{v c \boldsymbol{k}, v' c' \boldsymbol{k'}} A^{\nu}_{v' c' \boldsymbol{k'}} = E^{\nu} A^{\nu}_{v c \boldsymbol{k}} \, ,
\end{equation}
which introduces the Coulomb interaction by a new basis set $\{A^{\nu}_{v c \boldsymbol{k}} \}$.
Here, the summation runs over the different valence bands $v'$, conduction bands $c'$ and Bloch vectors $\boldsymbol{k'}$. The effective Hamiltonian $H^{e}$ contains the difference of the quasiparticle energies of electron and hole as well as the electron-hole interaction that is responsible for the formation of bound excitonic states.
$E^{\nu}$ denotes the energy of the excitonic state $\nu$ and $A^{\nu}_{v c \boldsymbol{k}}$ is the associated eigenvector for bound electron-hole states.
Selecting a spectral range of a specific excitonic transition, we focus on a two band system (determined by the detected frequency range $\omega$ and momentum transfer $\Delta\boldsymbol{q}$) with a conduction band ($c$) and a valence band ($v$) and rewrite the electronic operators in terms of pair-operators~\cite{katsch_theory_2018}:
\begin{equation}
\label{eq:introducing_pair-operators}
    \begin{aligned}
        c^{\dagger}_{\boldsymbol{k_1}} v_{\boldsymbol{k_2}}&=P^{\dagger}_{\boldsymbol{k_1},\boldsymbol{k_2}} 
        \\
        v^{\dagger}_{\boldsymbol{k_2}} c_{\boldsymbol{k_1}} &= P_{\boldsymbol{k_1},\boldsymbol{k_2}}
        \\
        c^{\dagger}_{\boldsymbol{k_1}} c_{\boldsymbol{k_2}} &= \sum_{\boldsymbol{k}} P^{\dagger}_{\boldsymbol{k_1},\boldsymbol{k}} P_{\boldsymbol{k_2},\boldsymbol{k}}
        \\
        v_{\boldsymbol{k_1}} v^{\dagger}_{\boldsymbol{k_2}} &= \sum_{\boldsymbol{k}} P^{\dagger}_{\boldsymbol{k},\boldsymbol{k_1}} P_{\boldsymbol{k},\boldsymbol{k_2}}.
    \end{aligned}
\end{equation}
This way, our approach is the following: From Eq.~\ref{eq:BSE_eigenvalue_equation}, we extract the two band model for Eq.~\ref{eq:introducing_pair-operators} and expand the pair operators into the wave function set $A^{\nu}_{v c \boldsymbol{k}}  \equiv \varphi^{\nu}_{\boldsymbol{k}}$. To obtain compact formulas, we separate the relative and center-of-mass (COM) motion of the two band exciton and introduce relative and COM  coordinates $\boldsymbol{Q}=\boldsymbol{k_1} - \boldsymbol{k_2}$, $\boldsymbol{\kappa}=\beta \boldsymbol{k_1}+\alpha\boldsymbol{k_2}$,  where $\alpha = \frac{m_e}{m_e+ m_h}$, $\beta = \frac{m_h}{m_e+m_h}$ and $m_e/m_h$ are the effective masses of electron or hole, to expand the pair-operators into an excitonic basis:
\begin{equation}
\begin{aligned}
P^{\dagger}_{\boldsymbol{k_1},\boldsymbol{k_2}} &= \sum_{\nu} \varphi^{*\,\nu}_{ \beta \boldsymbol{k_1}+\alpha\boldsymbol{k_2}} P^{\dagger\, \nu}_{\boldsymbol{k_1} - \boldsymbol{k_2}} \equiv \sum_{\nu} \varphi^{*\,\nu}_{ \boldsymbol{\kappa}} P^{\dagger\, \nu}_{\boldsymbol{Q}}
     \\
P_{\boldsymbol{k_1},\boldsymbol{k_2}} &= \sum_{\nu} \varphi^{\nu}_{ \beta \boldsymbol{k_1}+\alpha\boldsymbol{k_2}} P^{ \nu}_{\boldsymbol{k_1} - \boldsymbol{k_2}} \equiv \sum_{\nu} \varphi^{\nu}_{ \boldsymbol{\kappa}} P^{ \nu}_{\boldsymbol{Q}}
    \end{aligned}
\end{equation}
where the wave functions and the approximated dispersions are solutions to the BSE. 
\clearpage
The free time-evolution of the excitonic operators reads:
\begin{equation}
    P^{ \nu}_{\boldsymbol{Q}}(t) = P^{ \nu}_{\boldsymbol{Q}} e^{-i \Omega^{ \nu}_{\boldsymbol{Q}} t } 
\end{equation}
where $\hbar  \Omega^{ \nu}_{\boldsymbol{Q}} = E^{\nu} + \frac{\hbar^2 \boldsymbol{Q^2}}{2M} $ is the dispersion of the exciton. For the description of the excitonic part of inelastic scattering spectrum only the following contributions has to be considered{\color{violet}~\cite{sturm_theory_2026}}:
    \begin{equation}
\label{eq:time_correlation_electron_operators_localized_exciton}
\begin{aligned}
   & \braket{ a^{\dagger \lambda}_{\boldsymbol{k_1}}(t+\tau) a^{\lambda'}_{\boldsymbol{k_2}}(t+\tau)  a^{\dagger \mu}_{\boldsymbol{k_3}}(t)a^{\mu'}_{\boldsymbol{k_4}}(t) }
    \\=&            \braket{    v^{\dagger}_{\boldsymbol{k_1}} (t+\tau) c_{\boldsymbol{k_2}} (t+\tau) c^{\dagger}_{\boldsymbol{k_3}}(t)v_{\boldsymbol{k_4}}(t) }
\end{aligned}
\end{equation}
Inserting this into Eq.~\ref{eq:time-ordered-correlation-function ausgeschrieben} reads:
\begin{widetext}
\begin{equation}
\begin{aligned}
\label{eq:G_vccv_for_localized_excitons}
    G^{vccv}(\Delta\boldsymbol{q},t+\tau,t)= \left|\boldsymbol{A}^{\perp}_{X} \omega_{X}\frac{ r_{0}}{ R } N \right|^2 \left|\frac{V_{\text{UC}}}{a_B^3} \right|^2 &\sum_{\substack{\nu}} |M^{\nu}_{v,c}(\Delta\boldsymbol{q})|^2 e^{i(\omega_{X}-\Omega^{\nu}_{\Delta\boldsymbol{q}}) \tau},
    \end{aligned}
\end{equation}
\end{widetext}
where we introduced 
\begin{equation}
\label{eq:localized_exciton_matrix_element_inelastic_scattering}
M^{\nu}_{v,c}(\Delta\boldsymbol{q}) =\sum_{\boldsymbol{R}_{n}}  \varphi^{\nu}(\boldsymbol{R}_{n})  t^{v,c}_{\Delta\boldsymbol{q}}(\boldsymbol{R}_{n})e^{-i\beta \Delta\boldsymbol{q}  \cdot\boldsymbol{R}_{n}}    
\end{equation}
The scattering spectrum, Eq.~\ref{eq:definition_spectrum_with_time_ordered_correlation}, reads:
\begin{widetext}
    \begin{equation}
\label{eq:scattering_spectrum_localized_excitons}
    \begin{aligned}
        S(\Delta \boldsymbol{q},\omega) &= \left|\boldsymbol{A}^{\perp}_{X} \omega_{X}\frac{ r_{0}}{ R } N\right|^2 \left|\frac{V_{\text{UC}}}{a_B^3} \right|^2 \sum_{\substack{\nu}} \left|M^{\nu}_{v,c}(\Delta\boldsymbol{q})\right|^2 \frac{2\Gamma}{(\Gamma^2 + (\omega-\omega_{X}+\Omega^{\nu}_{\Delta\boldsymbol{q}})^2}
    \end{aligned}
\end{equation}
\end{widetext}
Eq.~\ref{eq:scattering_spectrum_localized_excitons} describes scattering from excitons and is determined by the Fourier transform of the excitonic wave functions $\varphi^{\nu}(\boldsymbol{R}_{n})$ at the lattice vectors $\boldsymbol{R}_{n}$, and the overlap integrals $t^{v,c}_{\Delta\boldsymbol{q}}(\boldsymbol{R}_{n})$, cf. Eq.~\ref{eq:localized_exciton_matrix_element_inelastic_scattering}. 
\\
So far, inelastic scattering from spatially delocalized Wannier excitons and spatially localized excitons is still accessible in Eq.~\ref{eq:G_vccv_for_localized_excitons} and Eq.~\ref{eq:localized_exciton_matrix_element_inelastic_scattering}. The main differences between both types of excitons occur in two different aspects of Eq.~\ref{eq:G_vccv_for_localized_excitons}:
Firstly, the ratio of unit cell volume and cubic excitonic Bohr radius, which for strongly localized excitons is of the order of unity $\frac{V_{\text{UC}}}{a_B^3}= \mathcal{O}(1)$ and thus differs significantly from delocalized Wannier excitons where typically $\frac{V_{\text{UC}}}{a_B^3}\ll 1$ (cf. Ref.~\cite{sturm_theory_2026}). Secondly, due to the spatial extent of the excitonic wave function in Eq.~\ref{eq:localized_exciton_matrix_element_inelastic_scattering}, the dominant contribution for localized excitons is the term $\boldsymbol{R}_{n}=\boldsymbol{0}$, whereas for Wannier excitons a broad range of $\{\boldsymbol{R}_{n}\}$ on the scale of the Wannier excitonic Bohr radius $a_{B}$ will contribute. Therefore, we split the contributions:
\begin{equation}
\label{eq:localized_exciton_matrix_element_split_into_dominant_and_perturbation}
\begin{aligned}
    M^{\nu}_{v,c}(\Delta\boldsymbol{q}) &= \varphi^{\nu}(\boldsymbol{0})  t^{v,c}_{\Delta\boldsymbol{q}}(\boldsymbol{0})
    \\
    &+\sum_{\boldsymbol{R}_{n}\neq \boldsymbol{0}}  \varphi^{\nu}(\boldsymbol{R}_{n})  t^{v,c}_{\Delta\boldsymbol{q}}(\boldsymbol{R}_{n})e^{-i\beta \Delta\boldsymbol{q}  \cdot\boldsymbol{R}_{n}}
    \end{aligned}
\end{equation}
and treat the sum over the non-vanishing lattice vectors $\boldsymbol{R}_{n}$ as a small perturbation when describing inelastic scattering from localized excitons. 
\\
\begin{widetext}
This provides a simple approximation for the spectrum of spatially \underline{loc}alized excitons:
\begin{equation}
\label{eq:scattering_spectrum_localized_excitons_with_assumptions}
    \begin{aligned}
        S_{\text{loc}}(\Delta \boldsymbol{q},\omega) &= \left|\boldsymbol{A}^{\perp}_{X} \omega_{X}\frac{ r_{0}}{ R } N\right|^2 \left|\frac{V_{\text{UC}}}{a_B^3} \right|^2 \sum_{\substack{\nu}} \left|\varphi^{\nu}_{\text{loc}}(\boldsymbol{0})  t^{v,c}_{\Delta\boldsymbol{q}}(\boldsymbol{0})\right|^2 \frac{2\Gamma}{(\Gamma^2 + (\omega-\omega_{X}+\Omega^{\nu}_{\Delta\boldsymbol{q}})^2}
    \end{aligned}
\end{equation}
This can be directly compared to the formula for spatially \underline{ext}ended Wannier excitons {\color{violet}~\cite{sturm_theory_2026}}:
\begin{equation}
\label{eq:inelastic_scattering_spectrum_wannier}
\begin{aligned}
    S_{\text{ext}}(\Delta\boldsymbol{q},\omega)&= \! \biggl|\boldsymbol{A}^{\! \perp}_{\! X} \omega_{X}\! \frac{ r_{0}}{ R }N\biggr|^2 \left|\frac{V_{\text{UC}}}{a_B^3} \right|^2 \sum_{\nu} \biggl| \varphi^{\nu}_{\text{ext}}(\boldsymbol{0}) \Delta\boldsymbol{q} \cdot \boldsymbol{r}_{v,c} \biggr|^2 \frac{2\Gamma}{(\Gamma^2 + (\omega-\omega_{X}+\Omega^{\nu}_{\Delta\boldsymbol{q}})^2}
    \end{aligned}
\end{equation}
\end{widetext}
where $\boldsymbol{r}_{v,c}=\braket{\lambda', \boldsymbol{0}| \boldsymbol{r} |\lambda, \boldsymbol{0}}$ is the position matrix element.
A direct comparison of the two limiting cases, Eq.~\ref{eq:scattering_spectrum_localized_excitons_with_assumptions} and Eq.~\ref{eq:inelastic_scattering_spectrum_wannier}, reveals that they can be obtained by replacing $t^{v,c}_{\Delta\boldsymbol{q}}(\boldsymbol{0}) \leftrightarrow \Delta\boldsymbol{q} \cdot \boldsymbol{r}_{v,c}$. While these quantities enter the scattering spectrum in an analogous manner, they address the interband transition ($v\to c$) from different perspectives: The transition element $t^{v,c}_{\Delta\boldsymbol{q}}(\boldsymbol{0})$ is determined by the spatial overlap of the electronic excitations of the conduction and valence band within the same unit cell, or more precisely by its Fourier transform with respect to the momentum transfer $\Delta\boldsymbol{q}$, cf. Eq.~\ref{eq:overlap_integral_definition_wannier_function}.
It therefore depends explicitly on the spatial structure of the Wannier functions $w_{\lambda}(\mathbf{r})$ and retains a non-trivial dependence on the momentum transfer. In contrast, the scattering spectrum of the spatially extended exciton Eq.~\ref{eq:inelastic_scattering_spectrum_wannier}, derived in Ref.~\cite{sturm_theory_2026}, is determined by the position matrix element $\boldsymbol{r}_{v,c}$. Due to the strong localization of the Wannier exciton in reciprocal space governed by the inverse excitonic Bohr radius ($a_{B}^{-1}$), $\boldsymbol{k}\cdot \boldsymbol{p}$ perturbation theory is applied to obtain the leading order, i.e. a linear dependency on the momentum transfer $\Delta\boldsymbol{q}$. In addition to the different interband transitions, both limiting cases obviously also differ in their respective wave function $\varphi^{\nu}$, the excitonic Bohr radii $a_{B}$ and the dispersion $\Omega^{\nu}_{\Delta\boldsymbol{q}}$. However, both spectra Eq.~\ref{eq:scattering_spectrum_localized_excitons_with_assumptions} and Eq.~\ref{eq:inelastic_scattering_spectrum_wannier} allow scattering only for $\varphi^{\nu}(\boldsymbol{0})$, which means that only optically active s-like excitonic states~\cite{kira_semiconductor_2011} contribute (in the respective limit) to the excitonic part of the inelastic scattering spectrum.
For the localized excitons, Eq.~\ref{eq:scattering_spectrum_localized_excitons_with_assumptions}, we assume that the exciton is spatially localized predominantly within a single unit cell, i.e. $\boldsymbol{R}_{n}=0$. For the Wannier exciton, Eq.~\ref{eq:inelastic_scattering_spectrum_wannier}, we assumed that only the lowest non-vanishing order in $\boldsymbol{k}\cdot\boldsymbol{p}$ perturbation theory is relevant. Both approximations obviously have to be checked individually for the considered material system.  
\\
In the following, we want to test the description of CT excitons with respect to experiments~\cite{abbamonte_dynamical_2008} and apply Eq.~\ref{eq:scattering_spectrum_localized_excitons_with_assumptions} to the alkali halide LiF.
\\

\section{Example: Exciton in LiF}
To verify our approach we calculate the scattering spectrum for LiF. This material has been extensively studied, which allows for a direct comparison to experimental data~\cite{abbamonte_dynamical_2008,roessler_electronic_1967}.
LiF is an insulator with a large, direct band gap at the $\Gamma$ point and shows a pronounced excitonic resonance in inelastic x-ray scattering experiments~\cite{abbamonte_dynamical_2008,balut_quantum_2025} which is energetically separated from other electronic resonances due to the large binding energy of the exciton~\cite{devreese_resonance_1972, rohlfing_electron-hole_1998, rohlfing_electron-hole_2000, abbamonte_dynamical_2008}. 
To evaluate the spectrum Eq.~\ref{eq:scattering_spectrum_localized_excitons}, the band structure and the associated wave functions of the electrons need to be determined, to directly provide the input for the evaluation of the BSE Eq.~\ref{eq:BSE_eigenvalue_equation} which determines the excitonic state. We calculate the band structure and the associated Bloch functions using \textsc{Quantum ESPRESSO} (QE)~\cite{giannozzi_quantum_2009,giannozzi_advanced_2017} within the local density approximation (LDA). 
The BSE formalism is implemented in \textsc{YAMBO}~\cite{marini_yambo_2009,sangalli_many-body_2019}, which is used together with the electronic-structure data obtained from QE to solve the BSE eigenvalue equation (Eq.~\ref{eq:BSE_eigenvalue_equation}). This provides the excitonic states $\varphi^{\nu}_{\boldsymbol{k}}$. Since \textsc{YAMBO} also allows the BSE to be solved for finite momentum transfer $\Delta\boldsymbol{q}$, the resulting eigenvalues can also 
be used to determine the exciton dispersion. We find that the assumed parabolic dispersion relation for the exciton is in good agreement with the first-principles results over a large portion of the Brillouin zone.
Deviations become apparent at the edges of the Brillouin zone where a parabolic dispersion relation would lead to discontinuities. We therefore fitted the first principles results with a cosine dispersion, which closely reproduces the parabolic behaviour for large parts of the Brillouin zone while ensuring a smooth dispersion at the edges. 
 The momentum transfer $\Delta q_{x}$ in the calculated spectra is chosen to be along the $(H,0,0)$ direction, where $(H,K,L)$ are Miller indices and is provided in units of $2\pi/a$, with the lattice constant of LiF $a=4.027\si{\angstrom}$~\cite{abbamonte_dynamical_2008}. To obtain the wave functions, required for the evaluation of Eq.~\ref{eq:scattering_spectrum_localized_excitons_with_assumptions}, we will compare an analytical tight-binding approach and an \textit{ab initio} approach where we use Wannier90~\cite{pizzi_wannier90_2020} to obtain \textit{maximally localized Wannier functions} (MLWFs) from the first-principles calculations performed in QE. 
\\
 Here, for the tight-binding approach, we construct the single-particle wavefunctions of valence and conduction band from localized atomic orbitals, which provides a simple and intuitive description of the underlying electronic structure. We also investigate to which extent each theoretical approach can reproduce the experimental data.
\subsection{Tight-binding approach}
To apply an analytical approach to Eq.~\ref{eq:scattering_spectrum_localized_excitons_with_assumptions}, we express the Wannier functions in terms of a linear combinations of atomic orbitals (LCAO):
\begin{equation}
    w_{\lambda}(\boldsymbol{r}-\boldsymbol{R}) = \sum_{n} a_n \phi_{n}(\boldsymbol{r}-\boldsymbol{R)}
\end{equation}
where $\phi_{n}$ represent the localized atomic orbitals and $a_n$ are the corresponding amplitudes. The conduction band in LiF is conventionally described by the Li-2s wave function whereas the degenerate valence bands are described by F-2p orbitals~\cite{abbamonte_dynamical_2008, wang_quasiparticle_2003}. To account for screening and shielding effects we apply Slater's rules~\cite{slater_atomic_1930} for the effective nuclear charges of Li and F respectively. The effective nuclear charges are given by $Z_{\text{eff}}^{\text{Li}}=1.3$ and $Z_{\text{eff}}^{\text{F}}=5.2$. The excitonic dispersion was interpolated using the results from the BSE calculations for different values of the momentum transfer $\Delta\boldsymbol{q}$.
\begin{figure}[ht!]
            \centering
         \includegraphics[width=0.5\textwidth]{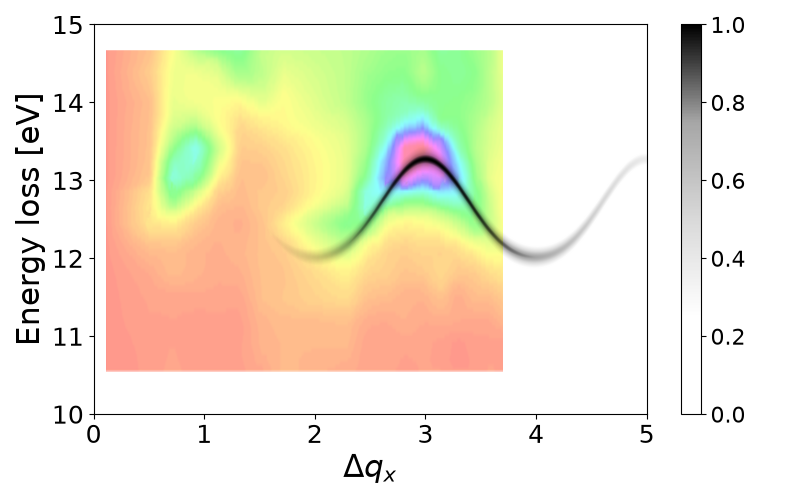}
    \caption{Normalized inelastic scattering spectrum of the exciton contribution in LiF using an analytical tight-binding approach. The adapted experimental data from Ref.~\cite{abbamonte_dynamical_2008}(Copyright (2008) National Academy of Sciences, U.S.A.) are shown in the background with the calculated spectrum overlaid for comparison.}
        \label{fig:scattering_spectrum_LiF}
\end{figure}
\\
Fig.~\ref{fig:scattering_spectrum_LiF} shows the contributions of the spatially localized exciton in LiF to the inelastic scattering spectrum Eq.~\ref{eq:scattering_spectrum_localized_excitons_with_assumptions} as a function of the energy loss $\hbar(\omega_{X}-\omega)$ and the momentum transfer $\Delta q_{x}$ in units of $2\pi/a$ in the $(H,0,0)$ direction. To provide a direct comparison of the theoretical spectrum Eq.~\ref{eq:scattering_spectrum_localized_excitons_with_assumptions} (black curve) to the experimental data (colored), we provided the latter in the background. The experimental results, which are adapted from  Ref.~\cite{abbamonte_dynamical_2008}, show the inelastic scattering spectrum in the spectral vicinity of the exciton for the momentum transfer $0\leq\Delta q_x \leq 3.77$. Here, two distinct excitonic contributions can be identified: a weaker feature centered around $\Delta q_x \approx0.8$ and a broader contribution with larger spectral weight that reaches its maximum at approximately $\Delta q_x =3$.
The numerically calculated spectrum is normalized to its highest value and in contrast to the experimental data, is shown for the momentum transfer $0\leq \Delta q_x \leq 5$. A broadening of $\hbar\Gamma = \SI{100}{\meV}$ is used and the exciton energy is shifted to the experimental values~\cite{abbamonte_dynamical_2008} to account for the underestimation of the band gap energy in the LDA approximation~\cite{deilmann_trions_2018, shirley_many-body_1998}.
The intensity of the spectrum increases with increasing momentum transfer $\Delta q_x$, starting at approximately $\Delta q_x=1.6$ until a maximum is reached at approximately $\Delta q_x=3$. Further increasing the momentum transfer results in a decreasing spectral weight, cf. Fig.~\ref{fig:scattering_spectrum_LiF}.
A comparison to the experimental data shown in Ref.~\cite{abbamonte_dynamical_2008} yields a good agreement in both, the exciton dispersion and the prediction of the maxima. Similar to the theory, the contributions of the experimental values vary between $1.8 \leq \Delta\boldsymbol{q} \leq 3.77$. In contrast to the experiment, we obtain no visible contribution for lower momentum transfer ($\Delta q_x\approx0.8$) and obtain contributions even for larger momentum transfer $\Delta q_x\geq 3.5$. The reason for this could be the simplified description of the conduction and valence band wave functions using the LCAO approach. In the following, we therefore want to compare the analytical approach with a first principles approach using the calculated band structure of QE as input for Wannier90~\cite{pizzi_wannier90_2020} to obtain MLWFs for the description of the valence and conduction bands in LiF.
\subsection{\textit{Ab initio} approach}
\begin{figure}[ht!]
            \centering
         \includegraphics[width=0.5\textwidth]{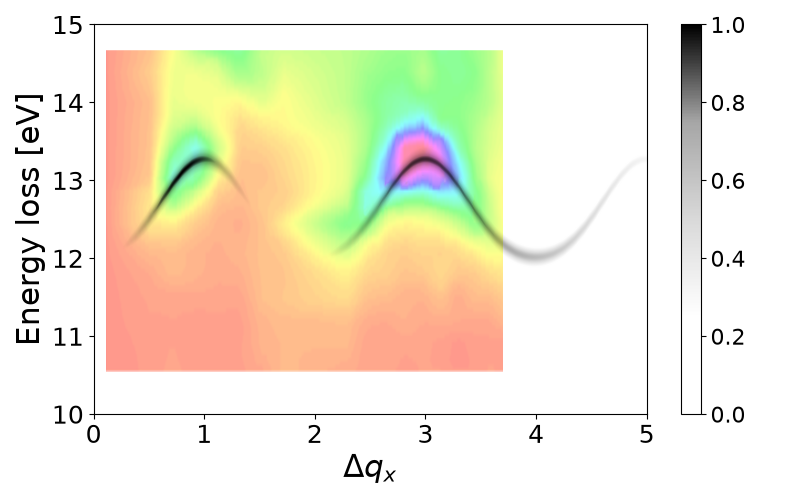}
    \caption{Normalized inelastic scattering spectrum of the exciton contribution in LiF using maximally localized Wannier functions. The adapted experimental data from Ref.~\cite{abbamonte_dynamical_2008}(Copyright (2008) National Academy of Sciences, U.S.A.) are shown in the background with the calculated spectrum overlaid for comparison.}
        \label{fig:scattering_spectrum_LiF_wannier90}
\end{figure}
Fig.~\ref{fig:scattering_spectrum_LiF_wannier90} shows the inelastic scattering spectrum which is calculated using MLWFs from first principles method that is in excellent agreement with other DFT-based theoretical results present in the literature~\cite{lee_first-principles_2013, gatti_exciton_2013}. Similar to Fig.~\ref{fig:scattering_spectrum_LiF} we show the experimental data (colored) in the background to provide a direct comparison to the theoretical results. It is evident, that by employing the MLWF approach, in contrast to the tight-binding approach (Fig.~\ref{fig:scattering_spectrum_LiF}) we now also obtain spectral features at lower momentum transfer centered around $\Delta q_x = 0.8$. A comparison to the experimental data reveals a slight overestimation of the spectral weight of this low momentum contribution and a larger extension towards both lower and higher momentum transfer of the theoretical inelastic scattering spectrum.
Similar to the LCAO approach, the spectral features for larger momentum transfer $\Delta q_x \geq 3.5$ are still obtained which are also present in other theoretical works~\cite{lee_first-principles_2013, gatti_exciton_2013}. 
The key spectral features of the experimental IXS~\cite{abbamonte_dynamical_2008} are therefore well obtained and significantly improved compared to the analytical LCAO approach, which underlines that a detailed knowledge of the respective wave functions is indeed essential for a qualitatively good description of the experiment. 
In other theoretically calculated spectra in the literature, the spectral weight for $\Delta q_x\leq 1.5$ is reduced by a factor of 2.6 to phenomenologically account for the screening of the plasmon in LiF~\cite{abbamonte_dynamical_2008, lee_first-principles_2013}. Although this assumption would increase the agreement with the experimental data in terms of the spectral weight, we decided to not include this in our work and instead provide the unadjusted theoretical results. 
\section{Conclusion}
We provided and tested a compact formula, Eq.~\ref{eq:scattering_spectrum_localized_excitons_with_assumptions}, for the theoretical description of excitonic contributions to the inelastic x-ray scattering spectrum. Throughout this work, we could demonstrate that, in particular, a comprehensive understanding of the wave function of electron and hole is essential for the accurate description of the localized excitonic inelastic scattering spectrum with respect to the momentum transfer. Nevertheless, we showed that even a simplistic approach, i.e. a LCAO approximation, already provides a qualitatively good agreement that contains the dominant spectral features of the experimental data~\cite{abbamonte_dynamical_2008}.. Although in the present study we focus on the description of localized excitons and provide an example for the inelastic scattering from the CT exciton in LiF, the approach is in general also applicable to spatially delocalized Wannier excitons~\cite{sturm_theory_2026}.
\newpage
\section{Acknowledgments}
J. S. and A. K. acknowledge financial support from the Deutsche Forschungsgemeinschaft (DFG) through Project No. 527838492.
\bibliographystyle{ieeetr}
\bibliography{references2}

\end{document}